\documentclass[a4paper]{article}

\usepackage[british]{babel} % set babel to british rather than american
\usepackage[utf8]{inputenc} % utf8 support in source code
\usepackage{lmodern}
\usepackage[T1]{fontenc} % better support for special characters in pdf but messes up title fonts, can be fixed by installing the debian package cm-super
\usepackage{textcomp}

\usepackage[a4paper,top=3cm,bottom=2cm,left=3cm,right=3cm,marginparwidth=1.75cm]{geometry}

\usepackage{authblk}

\usepackage{mathtools} % math packages
\usepackage{amsfonts}
\usepackage{amssymb}
\usepackage{physics}
\usepackage{tabu}

\usepackage{afterpage}

\usepackage{siunitx}
\usepackage{graphicx}
\graphicspath{{./Figures/}}
\usepackage{microtype}   
\usepackage{soul}

\usepackage[autostyle]{csquotes} % recommended by biblatex
\usepackage{xpatch} % recommended by biblatex
\usepackage[backend=biber, giveninits, sorting=none, style=numeric-comp]{biblatex} % much more flexible than BibTeX
\newcommand*{\m}{\mathrm}

\title{First Demonstration of a Pixelated Charge Readout for Single-Phase Liquid Argon Time Projection Chambers}

\author[3]{J.~Asaadi}
\author[1]{M.~Auger}
\author[1]{A.~Ereditato}
\author[1]{D.~Goeldi\thanks{Corresponding author: goeldi@protonmail.com}}
\author[1]{R.~H{\"a}nni}
\author[2]{U.~Kose}
\author[1]{I~Kreslo}
\author[1]{D.~Lorca}
\author[1]{M.~Luethi}
\author[1]{C.~Rudolf~von~Rohr}
\author[1]{J.~R.~Sinclair\thanks{Corresponding author: james.sinclair@lhep.unibe.ch}}
\author[1,2]{F.~Stocker}
\author[1]{C.~Tognina}
\author[1]{M.~Weber}

\affil[1]{Albert Einstein Center for Fundamental Physics, Laboratory for High Energy Physics,  University of Bern, 3012 Bern, Switzerland}
\affil[2]{CERN, 1211 Geneva, Switzerland}
\affil[3]{Department of Physics, The University of Texas at Arlington, Arlington, Texas 76019, USA}

\begin{document}
	\maketitle

\begin{abstract}
Liquid Argon Time Projection Chambers (LArTPCs) have been selected for the future long-baseline Deep Underground Neutrino Experiment (DUNE).
To allow LArTPCs to operate in the high-multiplicity near detector environment of DUNE, a new charge readout technology is required.  
Traditional charge readout technologies introduce intrinsic ambiguities, combined with a slow detector response, these ambiguities have limited the performance of LArTPCs, until now.
Here, we present a novel pixelated charge readout that enables the full 3D tracking capabilities of LArTPCs. 
We characterise the signal to noise ratio of charge readout chain, to be about 14, and demonstrate track reconstruction on 3D space points produced by the pixel readout.
This pixelated charge readout makes LArTPCs a viable option for the DUNE near detector complex.
\end{abstract}

\section{Introduction} \label{sec:Intro}

Liquid Argon Time Projection Chambers (LArTPCs) are ideal neutrino detectors due to their high density, homogeneous calorimetry, and the potential for precise 3D tracking.
Hence, LArTPCs have been selected as the far detector for the future long-baseline Deep Underground Neutrino Experiment (DUNE)~\cite{DUNE}.  
DUNE faces increasing sensitivity demands that will be met by high statistics and improved background rejection. 
To increase statistics at the far detector site, \SI{1300}{\kilo\metre} from the target, a neutrino beam $\order{\SI{1}{\mega\watt}}$ is required.
At the near detector, only \SI{574}{\metre} from the target, this beam intensity corresponds to $\order{0.1}$~events~per~tonne~per~beam~spill~\cite{DUNE2,DUNE3}.
To minimise detector response uncertainties between the near and far, it would be ideal to have a LArTPC component of the DUNE near detector complex.
Unfortunately, traditional LArTPCs are not suitable for near detector environments.

Since their evolution from gaseous TPCs~\cite{TPC,LArIonize,LArTPC}, the charge readout for LArTPCs has been achieved with two or more projective wire planes. 
Projective wire readouts have been successfully demonstrated in a number of experiments~\cite{icarus,argonute,uboner}, however they introduce intrinsic ambiguities in event reconstruction~\cite{ambiguous}. 
The ambiguities are due to reconstructing complex 3D shapes with a limited number of 2D projections, and are particularly problematic if tracks are aligned parallel to the wire plane, or multiple events overlap in drift direction.    
LArTPCs are slow detectors with a drift speed of \SI{2.1}{\milli\metre\per\micro\second} at \SI{1}{\kilo\volt\per\centi\metre}~\cite{protoLASER}, making event pile-up a problem for projective wire readouts in high-multiplicity near detector environments.
It is possible to increase voltages beyond \SI{1}{\kilo\volt\per\centi\metre}\cite{breakdown_16, latex}, however it is both safer and simpler to overcome pile-up with a charge readout free from ambiguities. 
For this reason, we have developed a novel approach based on a pixelated charge readout to exploit the full 3D potential of LArTPCs.

Pixelated charge readout is not a new idea, it has been employed in gaseous TPCs since the early 2000's~\cite{gaspix}. 
However, gaseous TPCs are less sensitive to power dissipation from readout electronics than single-phase LArTPCs. 
It is only relatively recently that cold readout electronics became available for LArTPCs~\cite{larasic}, with cold preamplifiers designed specifically for wire readouts.  
Existing wire readout electronics however cannot be applied to such a scheme due to the increase in channel number.  
Ideally, the charge collected at every pixel would be amplified and digitised individually.
To make use of existing cold wire readout electronics for the measurements described here, a form of analogue multiplexing had to be employed. 
While not ideal, this allowed for the proof of principle of a pixelated charge readout in a single-phase LArTPC.   
Bespoke pixel readout electronics are being developed~\cite{larpix} as a result of our work.

In this paper we demonstrate the primary goal of a pixelated charge readout, direct access to 3D space points for event reconstruction, and the characterisation of the Signal to Noise Ratio (SNR) of such a readout. 

\section{Experimental Set-up}

\subsection{Pixel PCB Design} \label{sec:PCB}
 
\begin{figure}[!ht]
\centering
\includegraphics[width=0.65\linewidth]{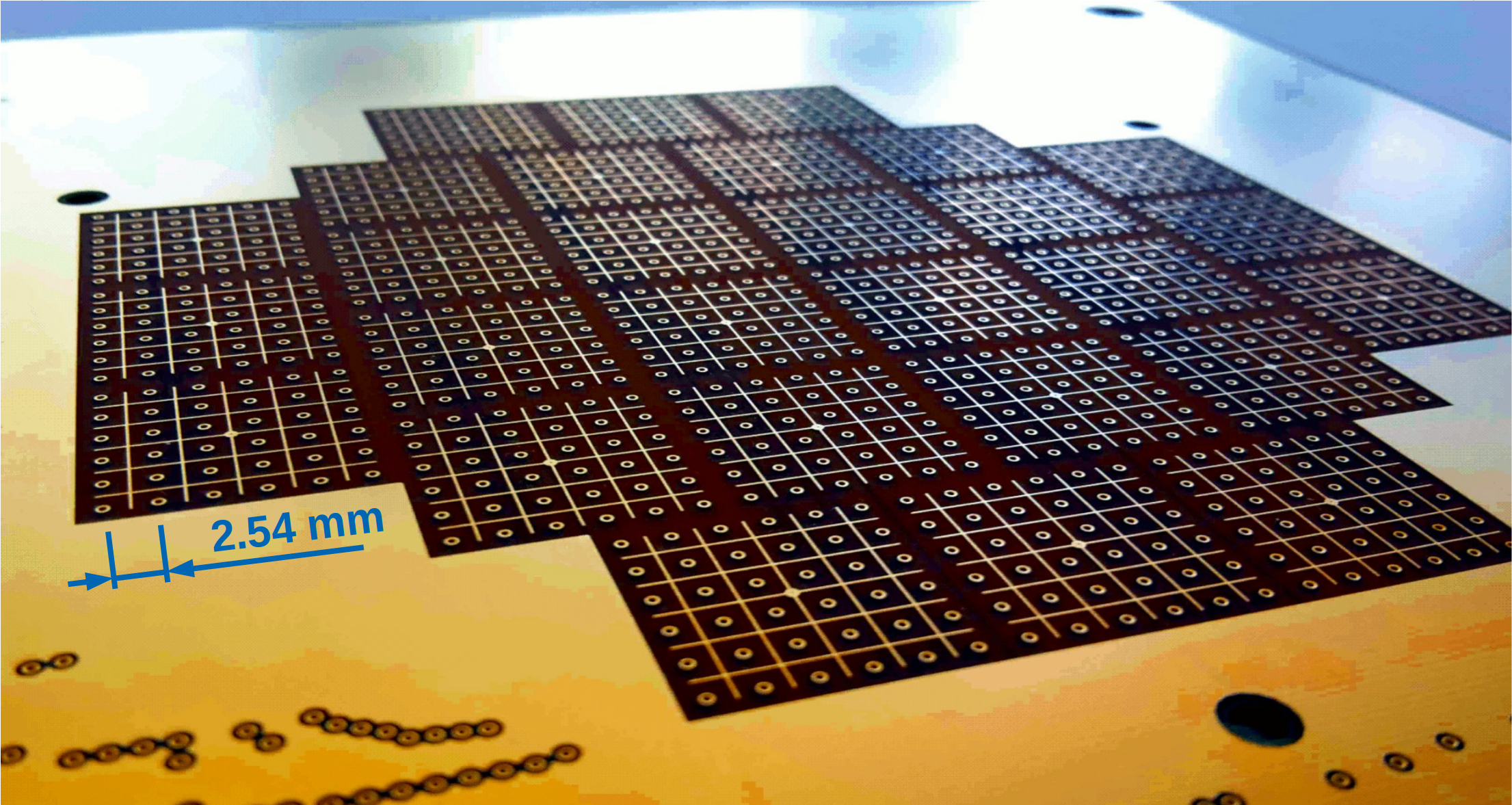}
\caption{Initial (July 2016) prototype pixelated anode PCB. The pixelated readout area is \SI{100}{\milli\metre} in diameter.
Each charge collection pixel is a \SI{900}{\micro\metre} via, at a pitch of \SI{2.54}{\milli\metre}, inductive focusing grids formed of \SI{152.4}{\micro\metre} copper traces surround the pixels. There are 28 inductive focusing grids with 36 pixels per region, a total of 1008 pixels.}
\label{fig:pixies}
\end{figure}

The pixelated anode plane used in our tests, shown in Figure~\ref{fig:pixies}, was produced as a conventional eight-layer Printed Circuit Board (PCB). 
The pixelated area is \SI{100}{\milli\metre} across, with pixels formed of \SI{900}{\micro\metre} diameter vias (PCB interlayer connections) with a pitch of \SI{2.54}{\milli\metre}.
An inductive focusing grid surrounds the pixels, it is made from \SI{152.4}{\micro\metre} wide copper traces split into 28 regions.
There are \num{6 x 6} pixels per region, giving a total of 1008 pixels. 

Vias were used for pixels instead of pads in order to minimise capacitance.
It is important that capacitance is minimised when amplifying charge since thermal noise scales with capacitance: $Q_{\mathrm{Noise}}=\sqrt{k_{\mathrm{B}}TC}$~\cite{noise}. 
To further minimise parasitic capacitance, the PCB design was optimised by removing unnecessary ground planes, routing signal tracks outside necessary ground planes, and increasing the thickness of the PCB to \SI{3.5}{\milli\metre} from an initial \SI{1.75}{\milli\metre}. 
Capacitance at each pixels is $\order{\SI{50}{\pico\farad}}$, however a significant contribution to this is due to additional traces required for the multiplexing scheme.

\begin{figure}[htb]
\centering
\includegraphics[width=0.65\linewidth]{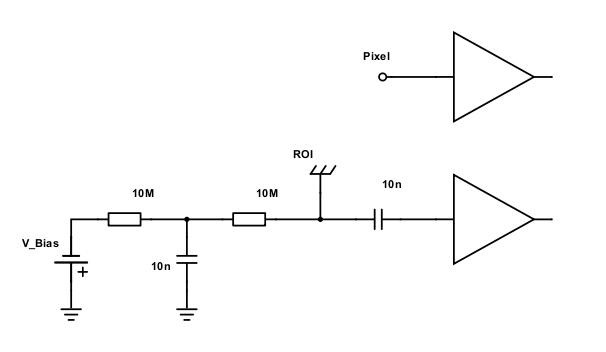}
\caption{Circuit diagram for pixel and inductive focusing grid (ROI). Pixels are directly coupled to the preamplifiers. ROIs are coupled to the preamplifiers via a \SI{10}{\nano\farad} capacitor, and are connected to the bias voltage via a \SI{10}{\mega\ohm} resistor and an RC filter. The RC filter consists of another \SI{10}{\mega\ohm} resistor and a \SI{10}{\nano\farad} capacitor to ground.}
\label{fig:circuit}
\end{figure}

As shown in Figure~\ref{fig:circuit}, the pixels are directly coupled to the preamplifiers.
The inductive focusing grids are coupled to the preamplifiers via a \SI{10}{\nano\farad} capacitor, and are connected to the bias voltage via a \SI{10}{\mega\ohm} resistor and an RC filter. 
The RC filter consists of another \SI{10}{\mega\ohm} resistor and a \SI{10}{\nano\farad} capacitor to ground.   

The bias on the inductive focusing grids had to be sufficient to allow full charge transparency (all charge collected by the pixels), yet low enough to minimise any risk of damaging the cold coupling capacitors.

\subsection{Readout Scheme}

Cryogenic preamplifiers were used to minimise both the noise-sensitive unamplified signal path and the thermal noise introduced by the amplifier itself~\cite{art_cold_ero}.
The preamplifier Application Specific Integrated Circuits (ASICs) used are the LARASIC4*~\cite{larasic} designed by the Brookhaven National Lab, first tested in the ARGONTUBE experiment~\cite{art_cold_ero} and deployed in the MicroBooNE and LArIAT experiments~\cite{uboner,lariat}.
LARASIC4*s were designed for traditional wire readouts, which require fewer channels than a pixelated readout of equivalent dimensions. 
Therefore, no focus was placed on high channel density, and the LARASIC4*s have only 16 channels per chip.
Given the 1008 pixels, cold digitisation is disfavoured due to power consumption constraints. 
Ideally, every pixel would be read out and the signals then digitally multiplexed, requiring bespoke pixel ASIC capable of cold amplification and digitisation for many channels.
Such ASICs are being developed by Lawrence Berkeley National Lab~\cite{larpix}, as a result of this work. 
Therefore, analogue multiplexing had to be used to minimise the channel numbers, with signals digitised at room temperature outside the cryostat.

The multiplexing scheme~\cite{maplesyrup} divides the pixels into a number of Regions Of Interest (ROIs).
Each ROI is defined as the pixels contained within a single inductive focusing grid.
All pixels at the same coordinate inside each of the ROIs are connected to the same DAQ channel, i.e. only one DAQ channel connects all the pixels in the top left corners of all ROIs, and so on. 
For a general expression of the multiplexing scheme, a pixel plane of $N \times N$ pixels (where $N = n ^ 2$ and $n$ integer) is divided into $n \times n$~ROIs, each ROI containing, again, $n \times n = N$ pixels.
Reading out such a plane requires $N$ DAQ channels for the ROIs plus another $N$ channels for the pixels.
With the employed multiplexing scheme, only as many pixel channels as there are pixels per ROI are required, due to the fact that all pixels at the same relative position across all ROIs share a common DAQ channel.
This means that an $N \times N$ pixel plane requires only $2 N$ DAQ channels; the same as a conventional 2-plane wire readout of the same pitch, and dimension.
Optimising the number of ROIs allowed us to readout the 1008 physical pixels with only 64 DAQ channels (28 ROIs + 36 pixels).

Pixel signals are then associated to an induction signal on the ROI grid as follows. 
If there is a signal on a certain pixel DAQ channel, the position inside the ROI is known but not which ROI.
By combining the inductive bipolar pulse on the ROI grid with any simultaneous collection pulses from the pixels, it is possible to disentangle the true position.
Again, the drawback of this approach is that it is not free from ambiguities; it fails for multiple simultaneous hits when it is impossible to say which pixel pulse belongs to which ROI pulse.
Ambiguous hits are flagged as pixel signals corresponding to multiple ROI signals, which can be disentangled later using reconstruction tools. 

\subsection{Pixel Demonstration TPC} \label{sec:viper}

\begin{figure}[!ht]
\centering
\includegraphics[width=0.9\linewidth]{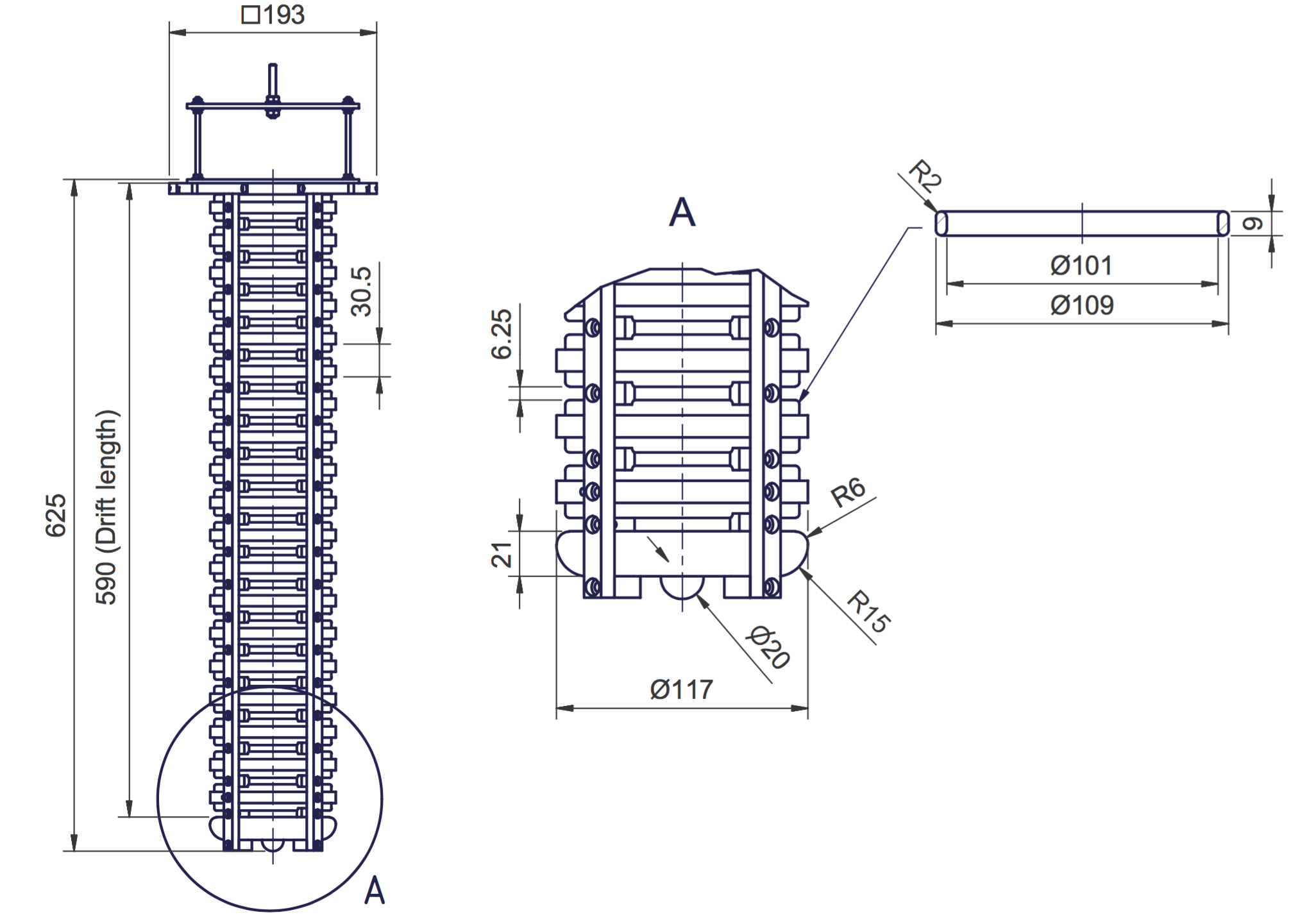}
\caption{\small Engineering drawing of the pixel demonstration TPC; \SI{590}{\milli\metre} drift length; \SI{6.25}{\milli\metre} field cage spacing; \SI{101}{\milli\metre} internal diameter.}
\label{fig:schematic}
\end{figure}

\begin{figure}[!ht]
\centering
\includegraphics[width=0.26\linewidth]{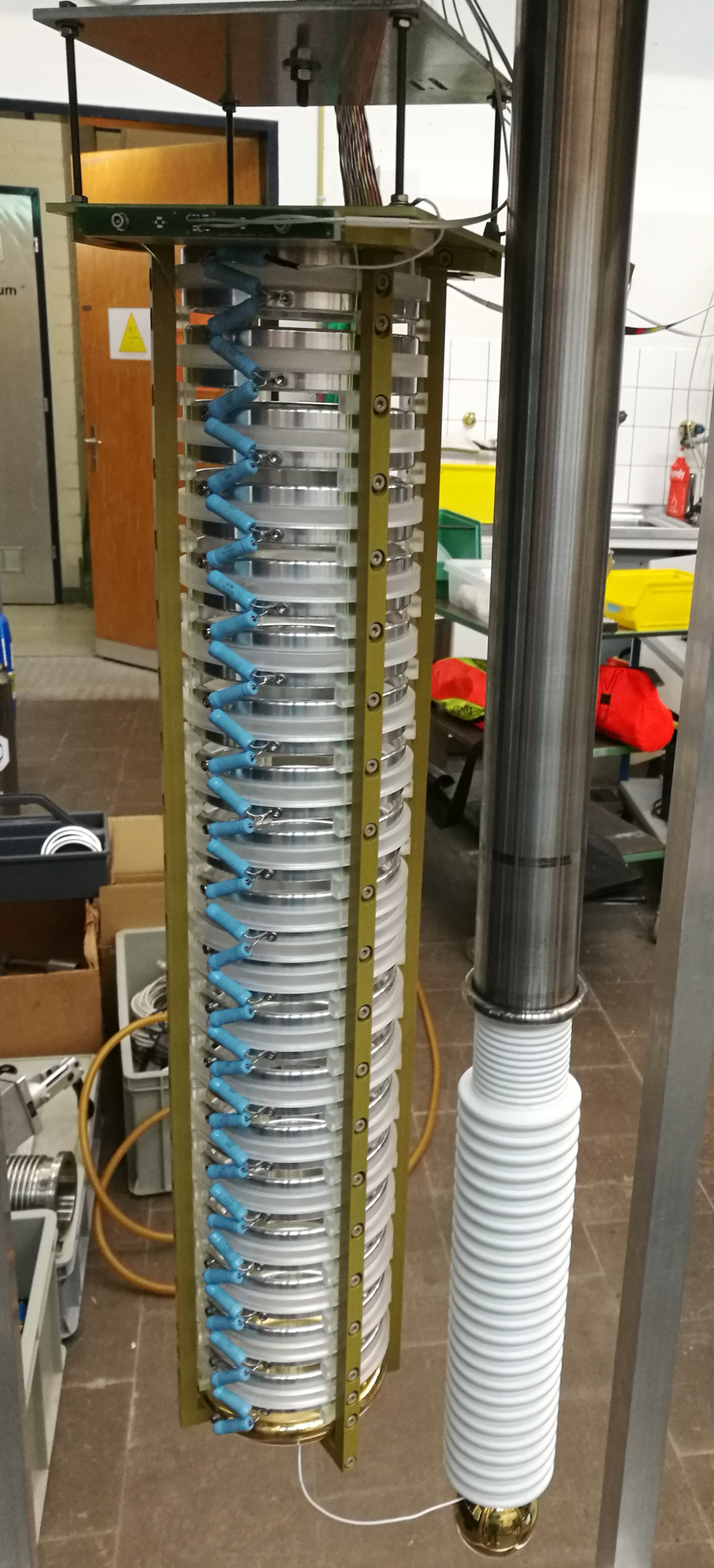}
\includegraphics[width=0.62\linewidth]{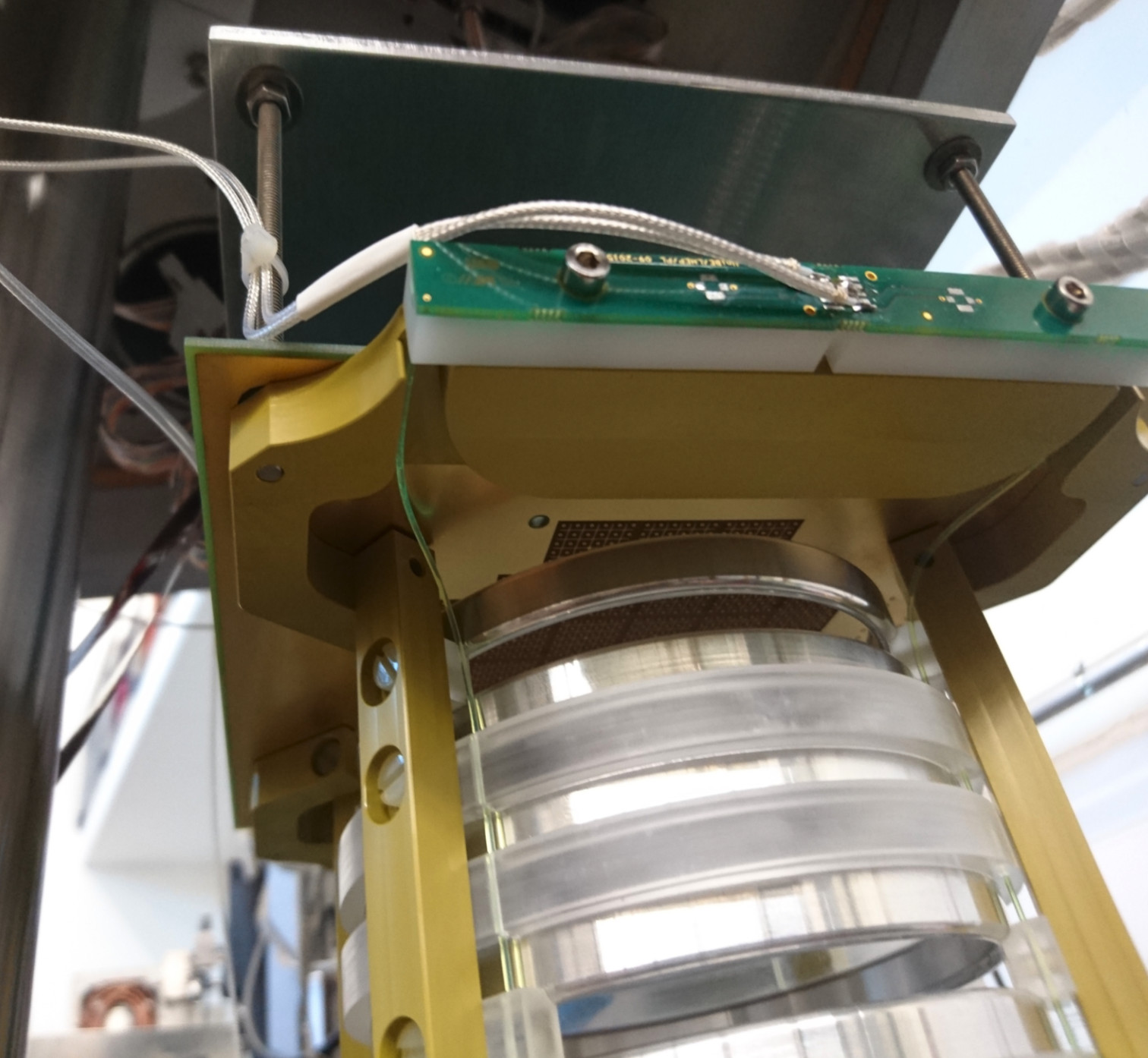}
\caption{\small  Left: Photograph of the  pixel demonstration TPC at Bern, with the HV feedthrough. Right: Close-up of the light collection system, showing wavelength shifting fibres coupling SiPMs to the TPB-coated light guides.}
\label{fig:v1per}
\end{figure}

The pixel demonstration TPC, shown in Figures~\ref{fig:schematic}~and~\ref{fig:v1per}, is cylindrical with an inner diameter of \SI{101}{\milli\metre} and a \SI{590}{\milli\metre} drift length. 
The TPC operated with a drift field of \SI{1}{\kilo\volt\per\centi\metre}, corresponding to a total drift time of \SI{281}{\micro\second}. 
The field-cage consists of aluminium rings supported by clear acrylic rings, with a cathode formed of a brass disc. 
The dimensions of the field-cage and cathode are shown in Figure~\ref{fig:schematic}.
Alternating acrylic rings are split, to allow for the circulation of purified LAr within the TPC volume.
Four square section PAI\footnote{Polyamide-imide} uprights support the cathode and field cage, with PEEK\footnote{Polyether ether ketone} screws fixing the pillars to the acrylic rings.
The four PAI uprights connect to a PAI frame which supports the anode plane and the light readout Silicon PhotoMultipliers (SiPMs), see Figure~\ref{fig:v1per}. 

The resistive divider consists of a chain of \SI{100}{\mega\ohm} Vishay Rox metal oxide resistors (ROX100100MFKEL). Each resistor is soldered to its neighbour, and fixed to the field cage at each joint with an M3 screw.   

The acrylic rings provide the light collection; their inner surfaces are machine-polished and coated with the WaveLength Shifter (WLS) TetraPhenylButadiene (TPB). 
The coating method is based on~\cite{TPBcoating}.
\SI{0.5}{\gram} of TPB and \SI{0.5}{\gram} of acrylic flakes were dissolved in \SI{50}{\milli\liter} of toluene and then mixed with \SI{12}{\milli\liter} of ethanol, which serves to increase the coating homogeneity. 
Three layers of the coating were applied by hand, with a fine brush. 

WLS fibres of \SI{1}{\milli\metre} diameter (Kuraray Y11(200)M) couple the acrylic rings to four SiPMs (Hamamatsu S12825-050P) mounted close to the anode, see Figure~\ref{fig:v1per}. 
The SiPMs and their front-end electronics were adapted from those developed at Bern for the cosmic ray taggers used in the MicroBooNE and SBND experiments~\cite{CRT, CRT2}.
For operation at LAr temperatures, the SiPM bias voltages had to be reduced from a nominal \SI{70}{\volt} at room temperature to \SI{53}{\volt}, in order to following the drop in breakdown voltage due to temperature.   
In the front-end electronics, two coincidences of two out of the four SiPMs are formed and combined by means of a logic \textit{OR} operation. 
This coincidence is used in order to improve trigger purity.

\subsection{Infrastructure}

The pixel demonstration TPC is housed in a double-bath vacuum-insulated cryostat with the outer bath open to atmosphere.
LAr is filtered first on filling through a pair of Oxysorb-Hydrosorb filters, and then recirculated through a single custom-made filter containing both activated copper and silica gel.
The cryostat and filtering method were previously used for LAr purity measurements~\cite{purititty}, and High-Voltage (HV) breakdown studies~\cite{HVoriginal}.
Based on these previous studies, an impurity concentration $\order{\SI{1}{ppb}}$ of oxygen-equivalent is estimated, which corresponds to a charge lifetime of \SI{290+-30}{\micro\second}.

The HV feedthrough remains unchanged from the breakdown studies; based on a PET-C polymer dielectric capable of conducting potentials as high as \SI{-130}{\kilo\volt}.
A low-pass filter was added between the power supply and feedthrough, which consists of an \SI{800}{\pico\farad} decoupling capacitor grounded between two \SI{100}{\mega\ohm} resistors connected in series, all of which is submerged in transformer oil.  

Only the warm signal path of the Data Acquisition (DAQ) was altered from that described in~\cite{protoLASER}, to include differential signalling.
An inverted waveform of the signal is put onto an additional conductor, and the difference is then taken between the two conductors.
Ground loops are avoided because the signal sink does not need connecting to the same ground as the signal source.
Additionally, the completely symmetric signal path means inductive noise pick-up is equal on both conductors and therefore cancelled at the signal sink.

\section{Data Analysis and Reconstruction} \label{sec:results}

The primary purpose of the experimental set-up described above is to demonstrate the principle of 3D
reconstruction utilising a pixelated charge readout within a LArTPC. Thus, in this section we focus on both
the characterization of the signal to noise ratio and the basic 3D track reconstruction that is
made directly possible by this technology. 
Since the purpose of the noise measurement is to characterise only the pixel readout, and not the whole TPC, both drift field and focusing bias were switched off. 
%For the same reason, the measured noise was compared to an expected signal as apposed to data.    

The HV for the TPC was set to \SI{-63}{\kilo\volt}, which, after a voltage drop across the HV filter and resistors, corresponds to a \SI{1}{\kilo\volt\per\centi\meter} drift field. The inductive focusing grid was set to a bias of \SI{-300}{\volt}, at which transparency was observed. 

\subsection{Signal to noise ratio}

To assess the Signal to Noise Ratio (SNR) dedicated noise data was taken employing a \SI{5}{\hertz} random trigger.
For the \num{2000} recorded events, all pixel and ROI channels were combined respectively and filled into amplitude distribution histograms.
The standard deviation of the two noise distributions was then calculated by fitting a Gaussian.
This value was used to calculate the noise for pixel and ROI channels according to
\begin{equation}
\m{SNR} = \frac{S}{\sigma}\,\m{,}
\label{eq:snr}
\end{equation}
where $\sigma$ is the noise standard deviation from the Gaussian fit and $S$ is the expected signal, which will be explained in detail below.
As can be seen in the left plot in Figure~\ref{fig:unfilteredRawData}, one of the pixel channels is significantly noisier in comparison to others, likely caused by a broken preamplifier.
Therefore, this channel was blinded for the SNR calculations.
The resulting equivalent noise charge is \SI{1095}{\elementarycharge} for the pixel channels and \SI{982}{\elementarycharge} for the inductive ROI channels.

The signal $S$ is often taken for a Minimum-Ionising Particle (MIP) as this is at the lower end of the signal range interesting for neutrino physics.
Getting a clean MIP sample from experimental data requires a calibrated reconstruction which was not available at the time of writing.
Therefore, we estimated the MIP signal from theory assuming an energy loss of \SI{2.1}{\mega\electronvolt\per\centi\metre}~\cite{pdg}.
This can be converted to charge loss using the energy required to produce one electron-ion pair: $W_{\m{i}} = \SI{23.6}{\electronvolt\per\elementarycharge}$~\cite{NobleGasDetectors}.
Additionally, charge recombination, diffusion and attachment losses characterised by lifetime need to be taken into account.
The recombination factor was measured by both the ICARUS and ArgoNeuT collaborations~\cite{icarusReco, argoneutReco}, and found to be $R_{\m{c}} \approx 0.7$ for a drift field of $\SI{1}{\kilo\volt\per\centi\meter}$.
For a non-zero drift field, diffusion needs to be split into longitudinal and transverse components.
Using the ARGONTUBE detector in Bern~\cite{argontube}, we measured a transverse diffusion coefficient $D_{\m{T}} = \SI{5.3}{\centi\metre\squared\per\second}$ at \SI{0.25}{\kilo\volt\per\centi\metre} while Gushchin et al.~\cite{gushchin} report a value of $D_{\m{T}} = \SI{13}{\centi\metre\squared\per\second}$ at \SI{1}{\kilo\volt\per\centi\metre}.
Even using the more conservative value, this results~\cite{lngDet} in a transverse spread of
\begin{equation}
\sigma_{\m{T}} = \sqrt{2 D_{\m{T}} t} \approx \SI{0.9}{\milli\metre}\,, 
\end{equation}
for our drift time of $t = \SI{281}{\micro\second}$; a value well below the pixel pitch of $d_{\m{p}} = \SI{2.54}{\milli\metre}$.
Considering that the longitudinal component is smaller than the transverse ~\cite{lngDet}, we neglect diffusion completely for our calculations.
Finally, our lifetime of \SI{290}{\micro\second} will result in the reduction of charge by a factor of $\approx\num{0.38}$ over the full drift distance.
Combining this, we get a signal of 
\begin{equation}
S = \dv{E}{x}_{\m{MIP}} \frac{R_{\m{c}} d_{\m{p}}}{W_{\m{i}}} = \SI{15821}{\elementarycharge}\,,
\end{equation}
for a charge deposited adjacent to the readout plane, and $S = \SI{6004}{\elementarycharge}$ for a charge deposited adjacent to the cathode.

Table~\ref{tab:snr} lists the SNR values obtained from these signal values and the aforementioned measured equivalent noise charge, using Equation~\eqref{eq:snr}.

\begin{table}[htb]
	\centering
	\caption{SNR values obtained from Equation~\eqref{eq:snr} using the theoretical signal of a MIP at the readout plane or cathode, respectively combined with the average equivalent noise charge for pixel and ROI channels obtained from measurements.}
	\label{tab:snr}
	\begin{tabu} to \textwidth {|l|l|S|}
		\hline
		{Channel} &	{MIP at} &			{SNR} \\
		\hline
		\hline
		{Pixel} &	{Readout plane} &	\num{14} \\
		\hline
		{Pixel} &	{Cathode} &			\num{5.5} \\
		\hline
		{ROI} &		{Readout plane} &	\num{16} \\
		\hline
		{ROI} &		{Cathode} &			\num{6.1} \\
		\hline
	\end{tabu}
\end{table}

\clearpage 

\subsection{3D track reconstruction}

A sample of several thousand cosmic ray events were collected, mostly minimum ionising muons traversing the TPC, to demonstrate 3D track reconstruction.
These events were triggered by the light readout described in Section~\ref{sec:viper}.

The reconstruction procedure comprises five steps:
\begin{enumerate}
	\item Noise filtering
	\item Hit finding
	\item Hit matching
	\item Ambiguity rejection
	\item Track fitting
\end{enumerate}

\begin{figure}[htb]
	\centering
	\includegraphics[width=.49\textwidth]{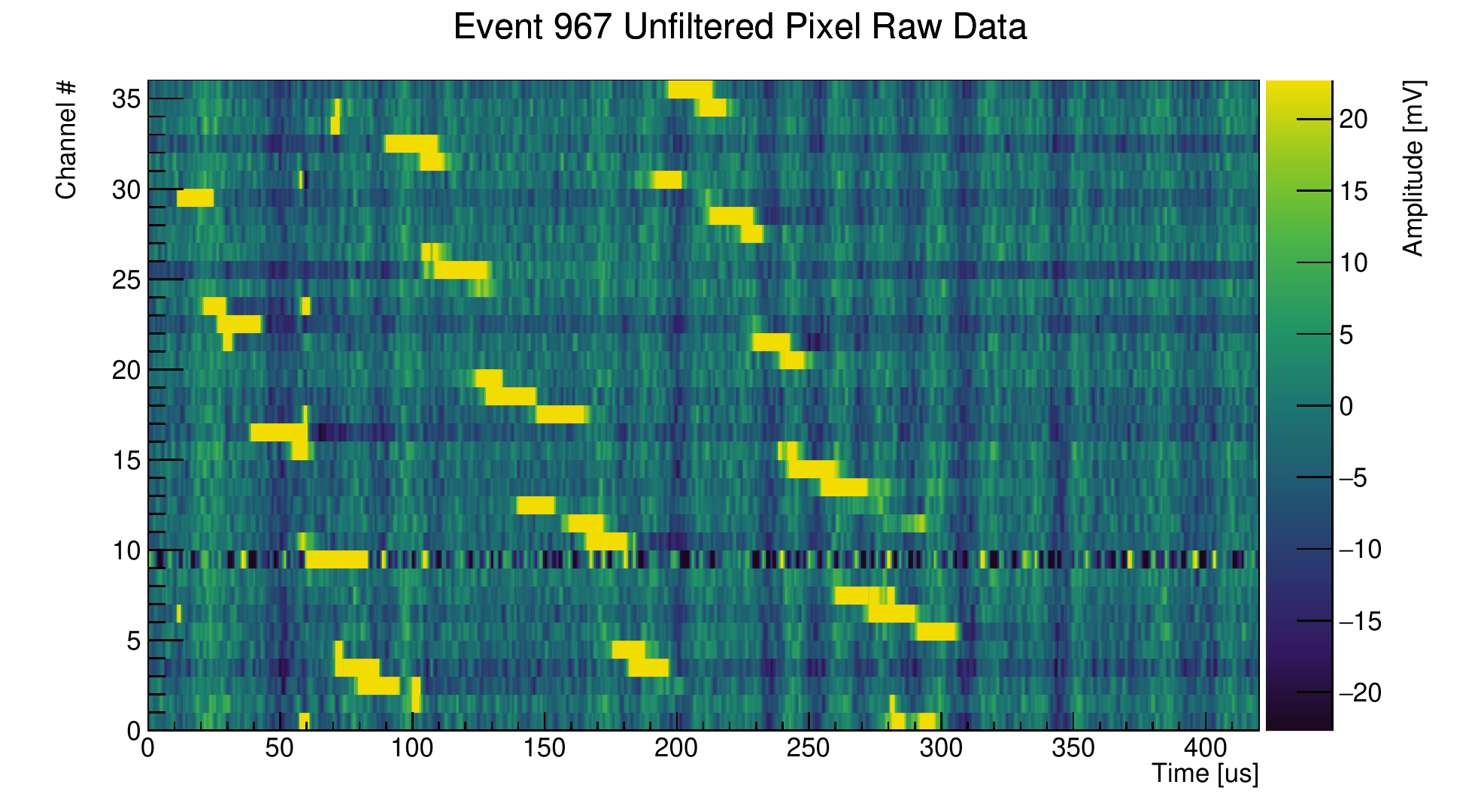}
	\includegraphics[width=.49\textwidth]{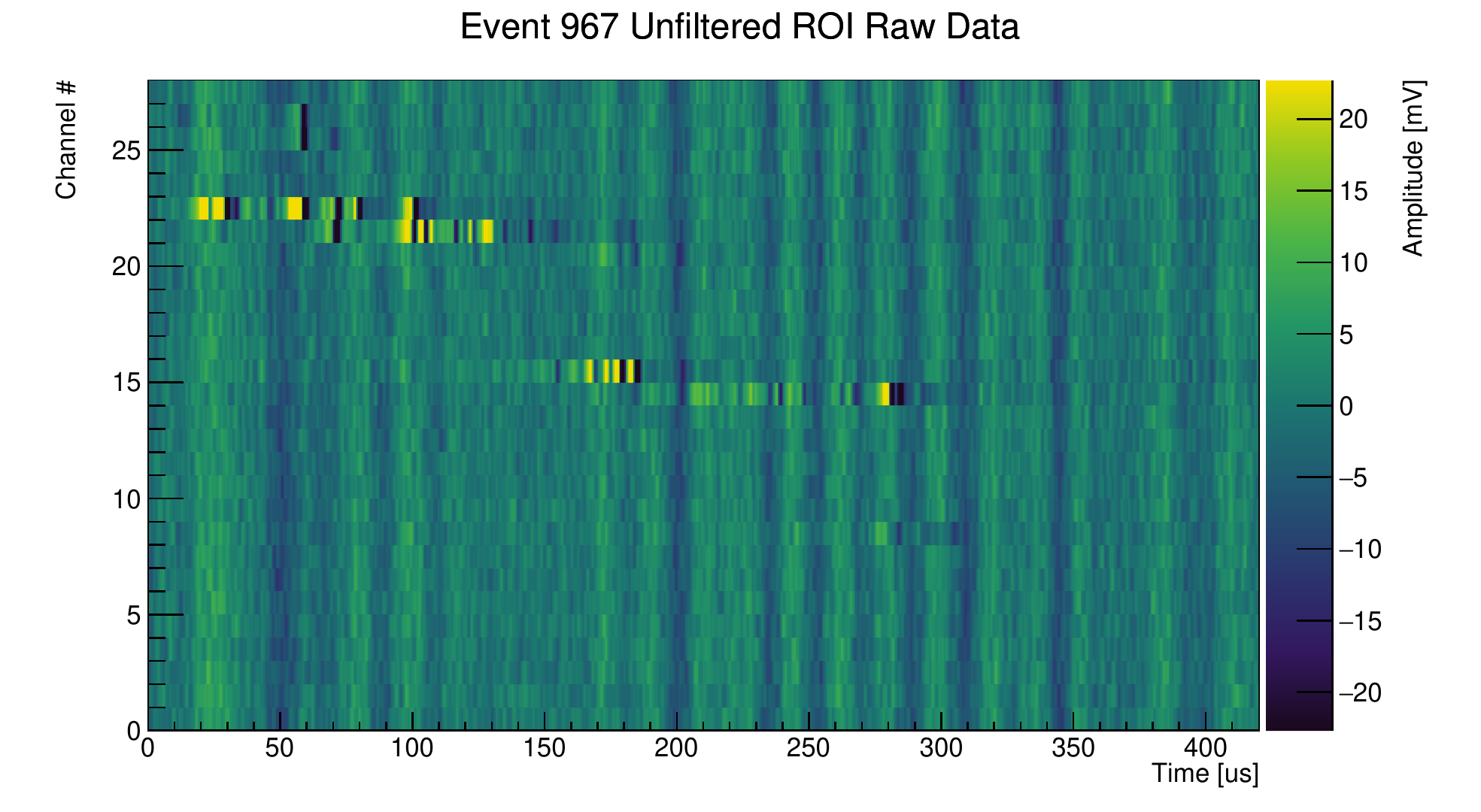}
	\caption{Unfiltered raw data of a typical MIP event (the same event as in Figures~\ref{fig:unfilteredRawData}~through~\ref{fig:kalman}).
		The left plot shows pixel data while the right plot shows ROI data.
		Note that the colour scale was adjusted to highlight the charge signals.
		Therefore, most signal peaks are above/below the maximum/minimum of the colour scale.
		The full range of a typical signal can be seen in Figure~\ref{fig:hitFinder}.}
	\label{fig:unfilteredRawData}
\end{figure}

\begin{figure}[htb]
	\centering
	\includegraphics[width=.49\textwidth]{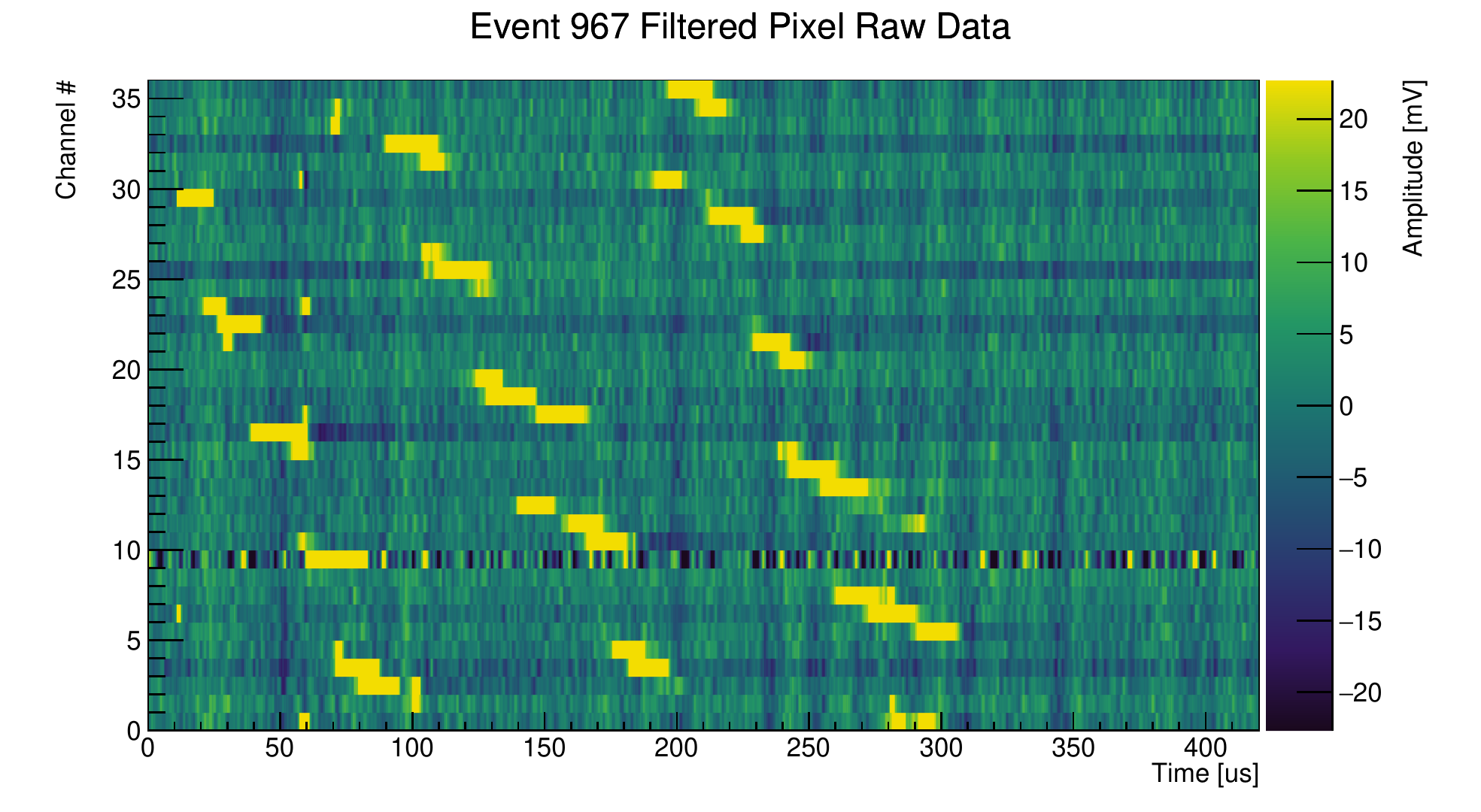}
	\includegraphics[width=.49\textwidth]{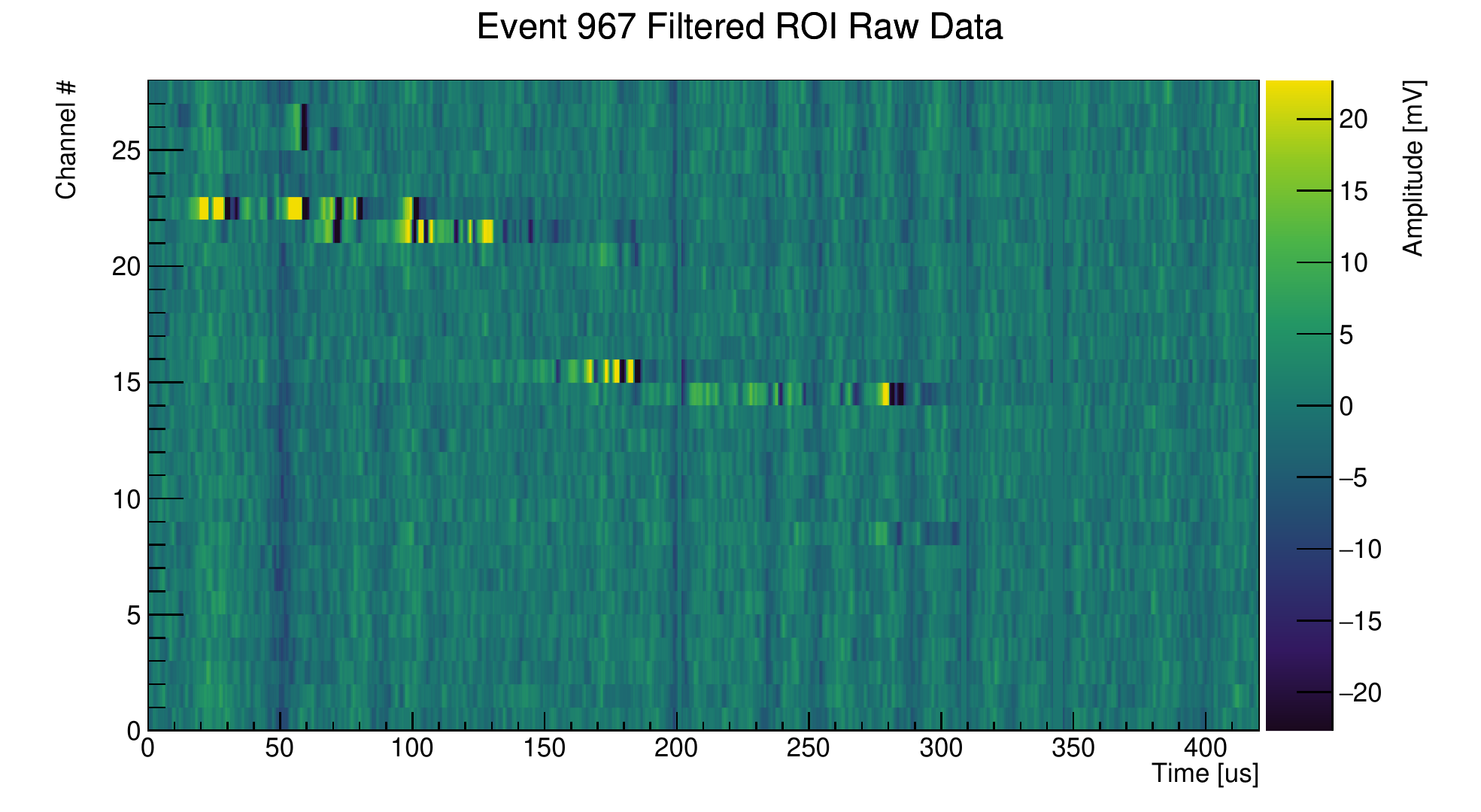}
	\caption{Filtered data of a typical MIP event (the same event as in Figures~\ref{fig:unfilteredRawData}~through~\ref{fig:kalman}).
		The left plot shows pixel data while the right plot shows ROI data.
		Note that the colour scale was adjusted to highlight the charge signals.
		Therefore, most signal peaks are above/below the maximum/minimum of the colour scale.
		The full range of a typical signal can be seen in Figure~\ref{fig:hitFinder}.}
	\label{fig:filteredRawData}
\end{figure}

\begin{figure}[htb]
	\centering
	\includegraphics[width=\textwidth, page=1]{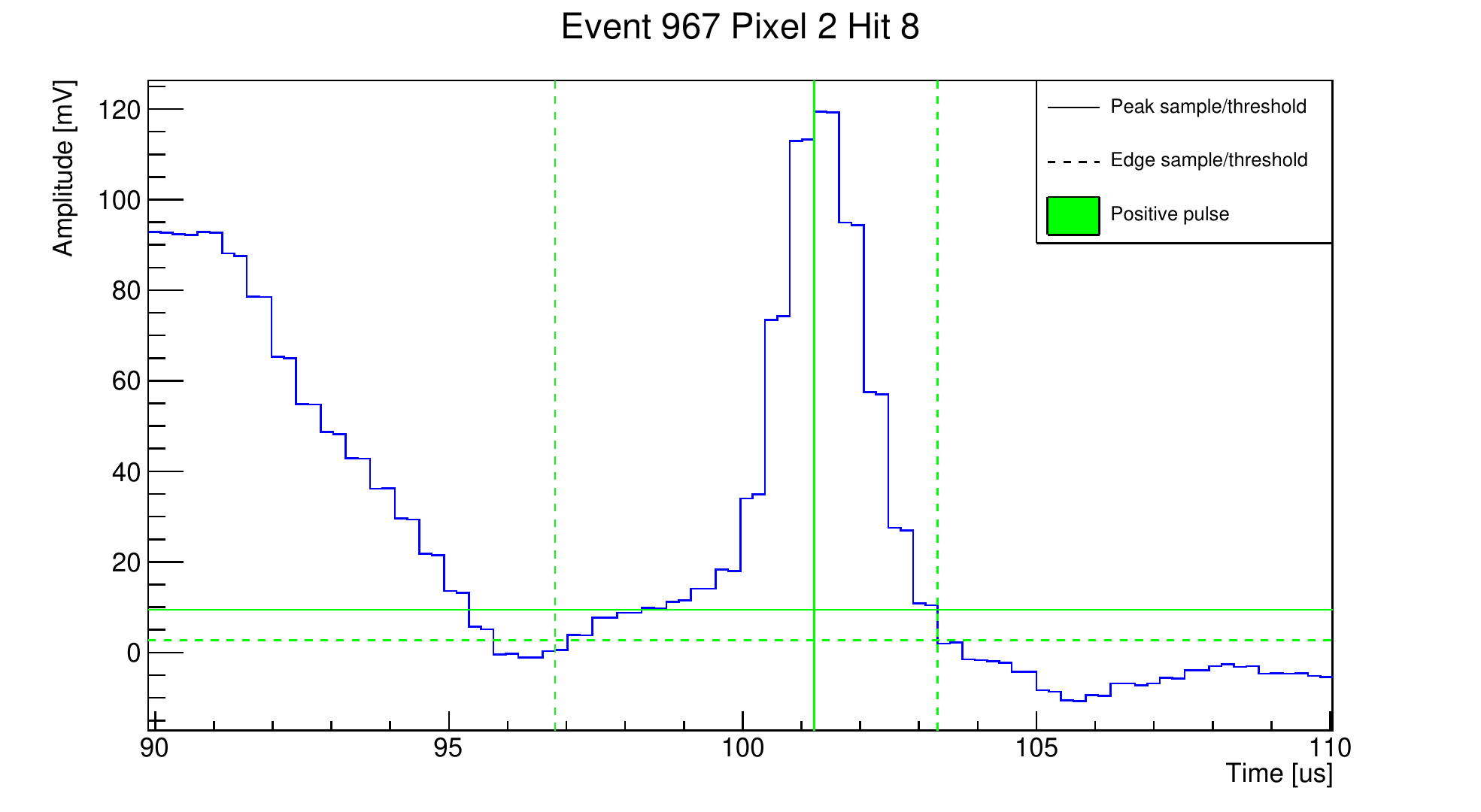}\\
	\includegraphics[width=\textwidth, page=3]{event967_pixel2_hit8}
	\caption{Pulse shapes of a single pixel (top) and ROI (bottom) hit of a typical MIP event (the same event as in Figures~\ref{fig:unfilteredRawData}~through~\ref{fig:kalman}).
		Superimposed are the thresholds of the hit finder algorithm. Horizontal lines represent thresholds: solid are the minimum thresholds required to be crossed for a pulse to be detected, and dashed are the thresholds used to detect the pulse edges.
		Vertical lines represent the corresponding detected peak/edge samples.
		Colour indicates a positive (green) or negative (red) pulse, or a zero crossing (yellow).}
	\label{fig:hitFinder}
\end{figure}

\begin{figure}[htb]
	\centering
	\includegraphics[viewport=600 0 1000 2000, clip, height=\textwidth, angle=90]{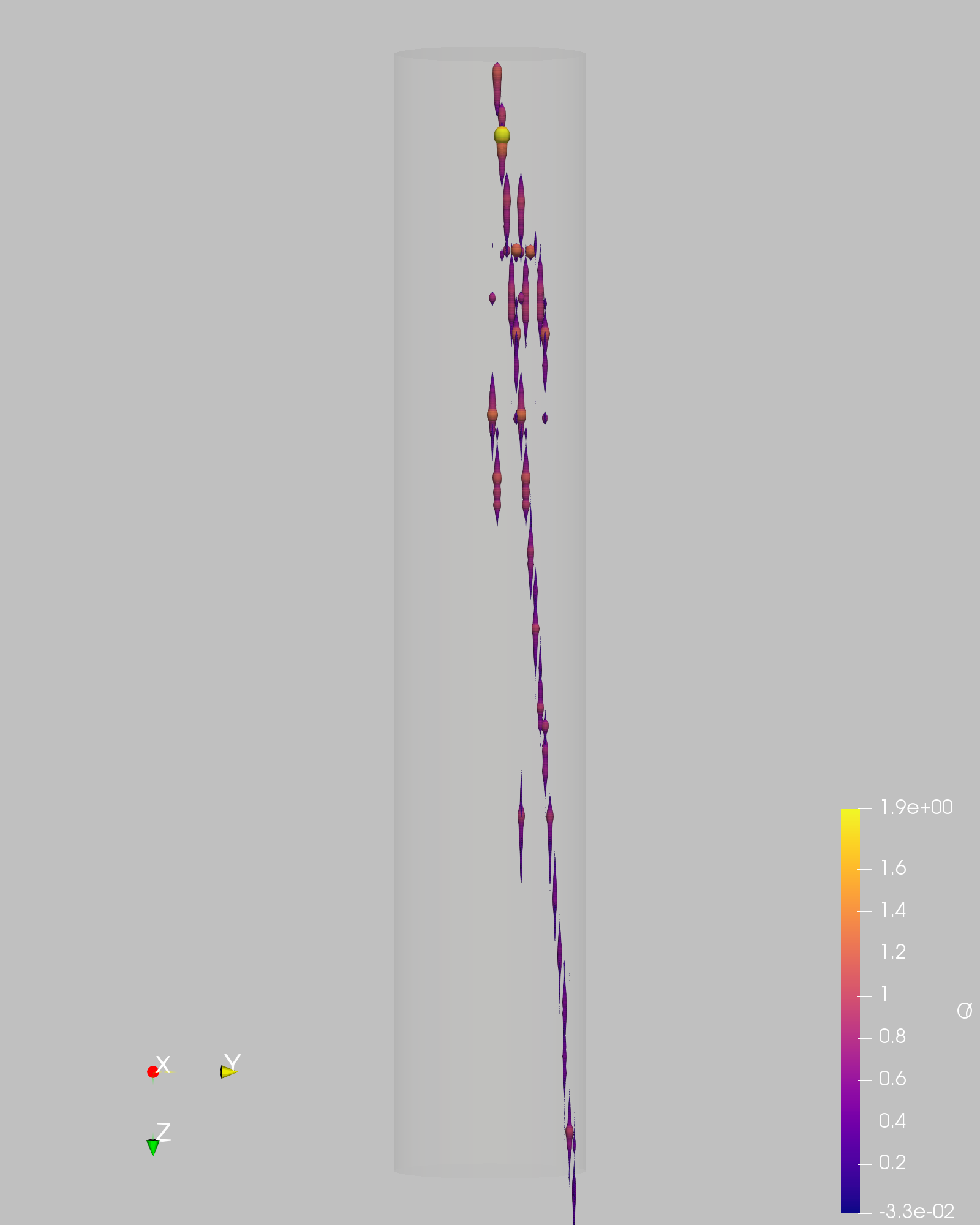} \\
	\includegraphics[viewport=600 0 1000 2000, clip, height=\textwidth, angle=90]{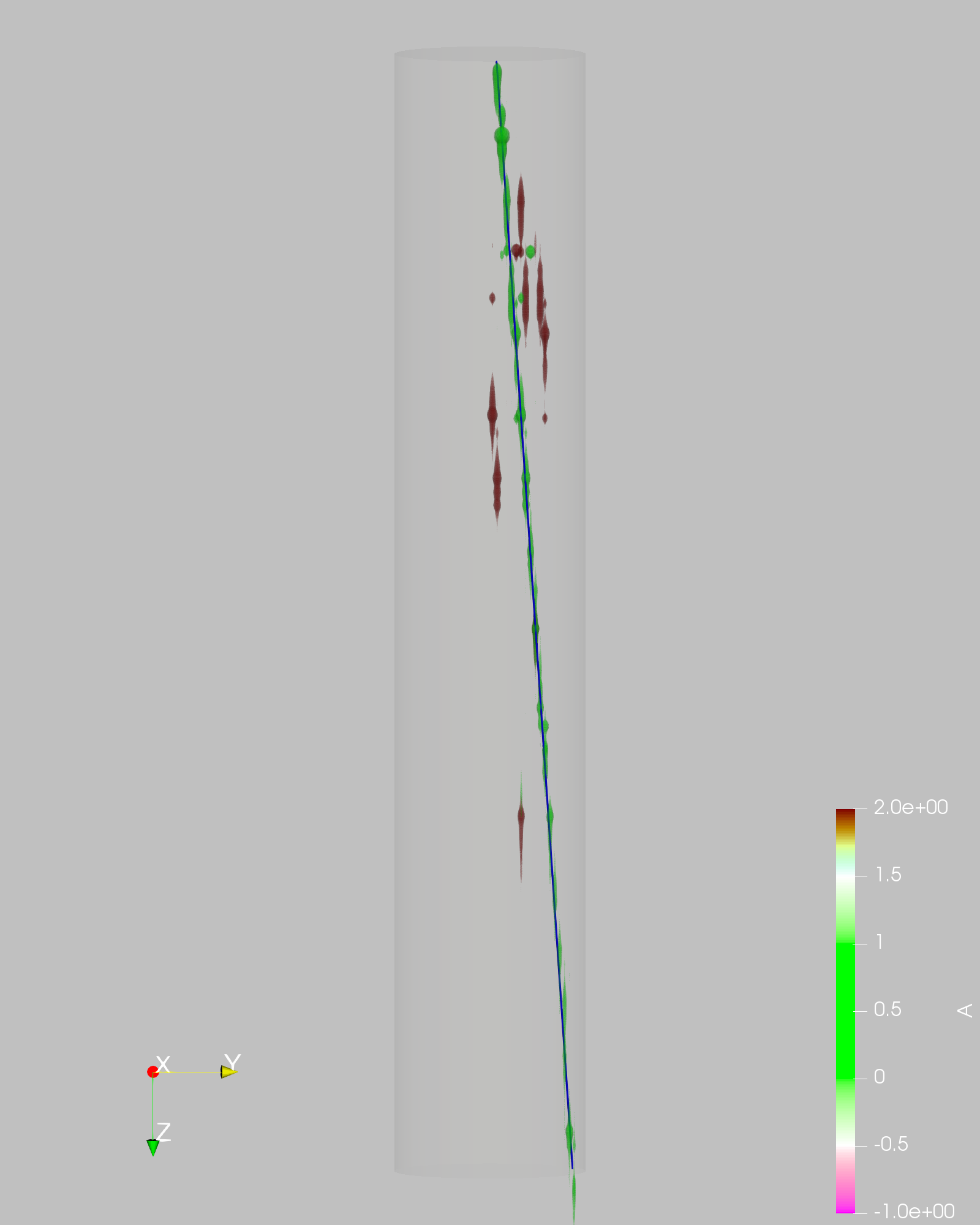} \\
	\includegraphics[viewport=600 0 1000 2000, clip, height=\textwidth, angle=90]{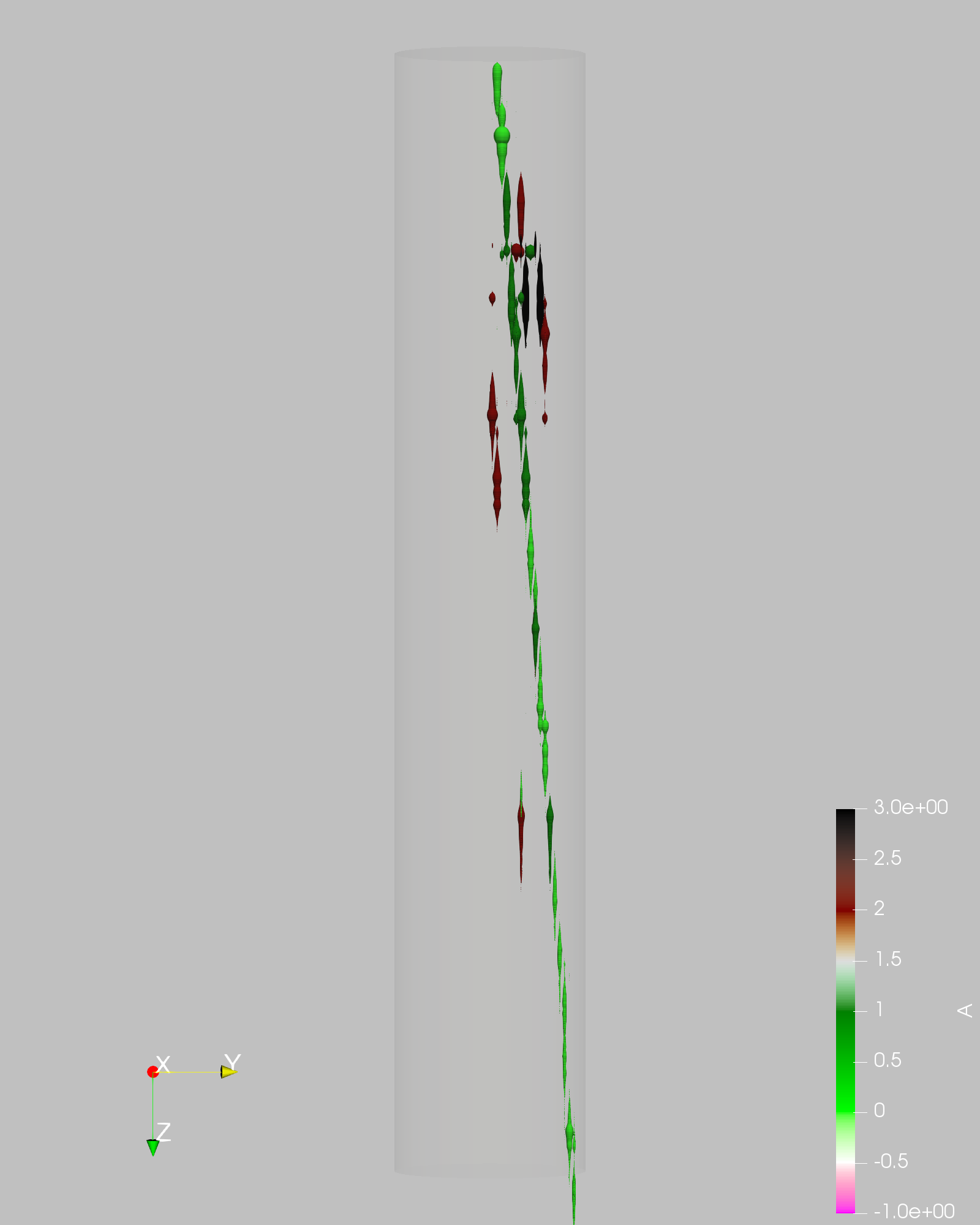}
	\caption{Reconstructed 3D hits from the hit finder.
		The passing particle is most likely a cosmic $\mu$ entering from the left (the same event as in Figures~\ref{fig:unfilteredRawData}~through~\ref{fig:kalman}).
		Drift direction is from right to left.
		Pulse shape is encoded as thickness.
		In the top plot, colour codes the amount of collected charge.
		The middle plot illustrates the ambiguity resolution employing a principal component analysis.
		Green hits are accepted while dark red ones are rejected.
		This is achieved by selecting the ambiguity closest to the eigenvector of the point cloud with the largest eigenvalue, represented by the blue line.
		In the bottom plot, the degree of ambiguity is colour-coded: Light green are unambiguous hits while dark green are selected solutions of ambiguous hits.
		Dark red through black are rejected solutions of ambiguous hits where darker colour represents a higher degree of ambiguity.
		As this is a quite clean track with only few short $\delta$ rays, there are no outliers rejected other than the multiplexing ambiguities.}
	\label{fig:pca}
\end{figure}

\begin{figure}[htb]
	\centering
	\includegraphics[width=\textwidth]{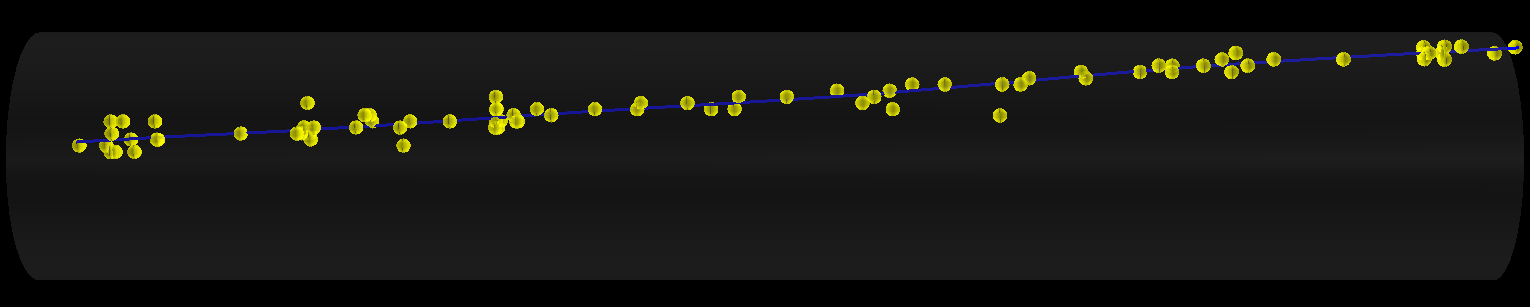}
	\caption{Track fitted by the Kalman filter.
		The TPC volume is shown in faint grey.
		The passing particle is most likely a cosmic $\mu$ entering from the left (the same event as in Figures~\ref{fig:unfilteredRawData}~through~\ref{fig:kalman}).
		Drift direction is from right to left.
		The yellow points are the input to the Kalman filter, the accepted hits from the principal components analysis.
		Blue is the output, a fitted track taking into account ionisation losses and multiple scattering in LAr.}
	\label{fig:kalman}
\end{figure}

These steps are explained in the following and depicted in Figures~\ref{fig:unfilteredRawData} through~\ref{fig:kalman}, all taken from the same MIP (cosmic muon) event.

In the first step, a noise-filtering algorithm is applied to the raw data.
As can be seen from Figure~\ref{fig:unfilteredRawData}, the noise is largely correlated across all the channels\footnote{Due to the much higher signal levels, the noise is barely visible on the pixel channels on the left.}.
This common-mode correlation can be exploited by the noise filter algorithm.
The following is done separately for the all pixel and ROI channels of each event.
Similarly to the SNR calculation, all samples are filled into an amplitude distribution histogram for each channel, and subsequently fitted with a Gaussian.
A noise band is defined per channel with its centre equal to the mean of the Gaussian and its width equal to the standard deviation multiplied by a tunable scaling factor.
The amplitudes of all channels within the corresponding noise band are then averaged for each sample.
Finally, this average is subtracted from each channel at the corresponding sample.
We chose this technique because it effectively suppresses the dominating common mode noise.
At the same time, spurious signals produced by high amplitudes from collected charge distorting the average are kept to a minimum by only accepting values within the noise band.
The effectiveness of the filtering can be seen in Figure~\ref{fig:filteredRawData}, showing the same raw data as Figure~\ref{fig:unfilteredRawData} post filtering.

The second step applies a recursive pulse finding algorithm.
The following is performed for each channel independently.
Most thresholds employed by the pulse finder are, again, defined in terms of noise amplitude.
Therefore, noise mean and standard deviation are recalculated after noise filtering.
A peak threshold is defined by multiplying the noise standard deviation by a variable scaling factor and adding the noise mean.
Then, the sample with the highest amplitude is found.
If it is below threshold, the process stops and proceeds to the next channel.
Otherwise, the pulse is scanned in positive and negative directions until it crosses respective lower noise thresholds.
After this, the whole pulse is stored and deleted from the input data then the process starts over with finding the new maximum sample and checking it against the peak threshold.
For stability reasons, the peak threshold relative to noise levels is compared against an absolute threshold and the higher of the two is applied.
The search is extended to the negative half-pulse for the bipolar ROI pulses.
The different thresholds employed and samples found by this process are illustrated in Figure~\ref{fig:hitFinder}.

Identified pulses are then combined into 3D hits by matching pixels pulses to ROI pulses.
For this proof of concept, this is done rather primitively by matching any pulses coinciding in time.
In Figure~\ref{fig:hitFinder}, a pixel and ROI pulse are matched if their time slices, defined by the vertical dashed lines, overlap.
This third step results in a rather high amount of ambiguities but assures that we do not miss any hits.

To resolve the ambiguities, a Principal Component Analysis (PCA) is applied to the 3D space points in a fourth step.
This technique is well established and described in literature, e.g.~\cite{pca}.
Therefore, we shall explain it only briefly here.
The basic idea is to calculate three orthogonal eigenvectors of the 3D space point cloud.
A graphic interpretation of these eigenvectors are the three axis of an ellipsoid fitted to the data points.
In case the points form a track, one of these eigenvectors will have a much higher eigenvalue than the other two.
This eigenvector is taken as an estimate for the track direction.
We resolve the ambiguities by selecting the one closest to the track estimate.
Furthermore, this procedure can be used to recursively reject outliers by forming a cylinder around the track estimate with a radius proportional to the second largest eigenvalue.
All hits outside this cylinder are rejected.
The procedure can be repeated by rerunning the PCA on the remaining points and performing another outlier rejection.
In a later stage of reconstructing more complex events, this algorithm can potentially be used to cluster 3D space points in order to separate multiple tracks.
The PCA ambiguity rejection is illustrated in Figure~\ref{fig:pca}.
It should be noted that readout electronics capable of cold amplification and digitisation, would render this step redundant. 

The final step consists of a Kalman filter for track identification.
For this, we used the well-established GENFIT track fitting package~\cite{genfit1, genfit2}.
Ionisation losses and multiple scattering are taken into account.
The particle is assumed to be a minimum-ionising muon with an initial momentum of \SI{260}{\mega\electronvolt} in the direction of the track estimate from the PCA.
We chose a recursive algorithm capable of dealing with outliers, a so-called \emph{deterministic annealing filter}.
This works by assigning successively lower weights to outliers with each recursion step.
For more details we refer to the respective publications~\cite{genfit1, genfit2}.
The resulting track is shown in Figure~\ref{fig:kalman}.

\afterpage{\clearpage}

In the near future, the Kalman filter will be capable of fitting the particle momentum and/or even particle type to the data.
However, at the time of this writing, this was not implemented.
In particular, the momentum stays roughly at the initial guess of \SI{260}{\mega\electronvolt}, assuming a minimum ionising muon in liquid argon.
A potential explanation for this is that the resolution of our detector is too low to estimate momentum from multiple scattering.
Another explanation might be the hit finder missing hits due to non-optimal tuning.
Proper tuning of the reconstruction requires a full simulation chain of the detector which is not yet available.
Using data to tune the reconstruction is prone to the introduction of circular biases.
On the other hand, most of the difficulties emerge from the multiplexing ambiguities and their resolution.
While the presented almost full 3D readout has already reduced the reconstruction complexity compared to a classical wire readout, an ambiguity-free readout will make reconstruction another big step easier by completely eliminating the need to resolve ambiguities.

\clearpage

\section{Summary} \label{sec:Summary}

We have presented a proof of concept for a pixelated charge readout for single-phase LArTPCs by successfully reconstructing 3D tracks of cosmic muons recorded by a prototype.
The requirement of high readout channel number has not yet been addressed.
In this first implementation, we have used existing wire readout electronics in conjunction with analogue multiplexing which introduces ambiguities.
Although much improved compared to classical wire readouts, the multiplexing ambiguities complicated reconstruction.
This work shows that it is of paramount importance to be capable of digitising the charge signals at cryogenic temperatures allowing for digital multiplexing and thus enabling a true, ambiguity-free, 3D LArTPC charge readout.
Work is currently under way to develop bespoke pixel readout electronics, based on the requirements highlighted by this demonstration. 
Once this last remaining problem is solved, pixelated charge readouts will enable the true 3D tracking capabilities of single-phase LArTPCs. 
This technology will provide the necessary reconstruction efficiency and background rejection to enable LArTPCs to operate in high-multiplicity environments, such as the DUNE near detector.

\section*{Acknowledgments}We acknowledge financial support of the Swiss National Science Foundation.
We would like to thank Yun-Tse Tsai and Tracy L. Usher of SLAC National Accelerator Laboratory, California, USA, for their valuable advice and guidance in the development of the 3D event reconstruction. 
We would like to thank Dean Shooltz of Michigan State University, Michigan, USA, for his extensive help in the design of the differential amplifiers.

\printbibliography

\end{document}